\documentclass{ws-procs9x6}

\begin{document}
\title{Electroweak Constraints \\on\\ Little Higgs Models\footnote{\uppercase{T}alk presented
at {\it \uppercase{SUSY} 2003:
\uppercase{S}upersymmetry in the \uppercase{D}esert}\/,
held at the \uppercase{U}niversity of \uppercase{A}rizona,
\uppercase{T}ucson, \uppercase{AZ}, \uppercase{J}une 5-10, 2003.
\uppercase{T}o appear in the \uppercase{P}roceedings.}}

\author{Patrick Meade}

\address{F.R. Newman Laboratory for Elementary-Particle Physics\\
Cornell University\\
Ithaca, NY 14853 USA\\
E-mail: meade@mail.lepp.cornell.edu}


\maketitle

\abstracts{In this talk I will give a brief introduction to Little
Higgs models in general, including an overview of all models in
existence thus far. I then review some of the generic constraints
on these models from electroweak precision measurements. }

\section{Motivation for Little Higgs}

The Standard Model~(SM) of particle physics is an incredibly good
effective field theory for particle physics below an energy of
around a TeV.  However, the Higgs mass in the SM is quadratically
sensitive to the cutoff of the SM.  This sensitivity manifests
itself through loop contributions to the Higgs mass involving the
top quark, gauge bosons, and the Higgs itself.  Experimental
evidence indicates that a light Higgs is preferred so something
must be done to obtain a light Higgs mass despite these quadratic
divergences.  The contribution to the Higgs mass can be written as
\begin{equation}
m_h^2=m_0^2-\mathcal{O}(1)\Lambda^2,
\end{equation}
where $m_h$ is the Higgs mass, $m_0$ is the bare mass and, and
$\Lambda$ is the cutoff of the SM.  If one desires a light Higgs
mass of a few hundred GeV, then if the cutoff of the SM is much
higher than a TeV one has to significantly fine tune the bare
Higgs mass against the quadratically divergent contributions.
Another approach to getting a light Higgs is to cancel the
quadratic divergences with physics beyond the SM.  Until recently
the only type of physics beyond the SM that is weakly coupled and
known to cancel the quadratic divergences was Supersymmetry.
Recently a promising new idea has arisen called Little Higgs
theories.

\section{What makes a Little Higgs?}

The basic idea for Little Higgs theories goes back to a much
earlier idea of having the Higgs be a pseudo-Goldstone
boson~(PGB)~\cite{georgi}.  If the Higgs were an exact Goldstone
boson~(GB) the Higgs would remain massless.  However, it would
only couple to other particles derivatively which is not how the
Higgs must couple to SM fields. Therefore one must introduce
couplings that make the Higgs a PGB so as to accommodate the
structure of the SM.  If one introduces these couplings naively
then the radiative corrections they introduce to the Higgs mass
are simply proportional to these couplings.  Since the couplings
in the SM such as the top Yukawa coupling are large this
reintroduces a fine-tuning to the Higgs mass.
\par  What makes a Little Higgs model is adding the crucial
new ingredient of ``collective symmetry breaking"~\cite{cohen} to
the PGB idea. Collective symmetry breaking is the idea that the
Higgs transforms under more than one symmetry and under each
individual symmetry the Higgs is an exact GB.  To break all the
symmetries you need at least two couplings, therefore at 1-loop
there are no quadratic divergences~(they do appear at higher loop
order).
\begin{table}[ht]
\tbl{Little Higgs models in existence thus far.  The models are
categorized by their global symmetries, gauge symmetries, whether
or not there is a triplet Higgs, and the number of light Higgs
doublets.} {\footnotesize
\begin{tabular}{|c|c|c|c|c|c|}
\hline Global Symmetries & Gauge Symmetries  &  triplet & \# Higgs
& ref \\ \hline
 $SU(5)/SO(5)$ & $[SU(2)\times U(1)]^2$ & Yes & $1$ & \cite{littlest} \\
 $SU(3)^8/SU(3)^4$ & $SU(3)\times SU(2)\times U(1)$ & Yes & $2$ & \cite{moose}\\
 $SU(6)/Sp(6)$ & $[SU(2)\times U(1)]^2$ & No & $2$ & \cite{anti}\\
 $SU(4)^4/SU(3)^4$& $SU(4)\times U(1)$ & No & $2$ & \cite{simple}\\
 $SO(5)^8/SO(5)^4$& $SO(5)\times SU(2)\times U(1)$ & Yes & $2$ & \cite{custmoose}\\
 $SU(9)/SU(8)$& $SU(3)\times U(1)$ & No & $2$ & \cite{simple2}\\
 $SO(9)/[SO(5)\times SO(4)]$& $SU(2)^3\times U(1)$ & Yes & $1$ & \cite{cust2} \\ \hline
\end{tabular}\label{models}
}
\end{table}
\par Even though there are a number of different Little Higgs models
there are still some generic features that all models have in
common. There is some global symmetry structure that is broken at
a scale $f$ to obtain the PGB Higgs. At around the scale $f$ there
will be new heavy gauge bosons, new heavy fermions, and some sort
of heavy triplet or singlet Higgs. In all Little Higgs models
there are still logarithmic divergences to the Higgs mass from the
heavy particles.  Because the mass of the heavy particles is
$\mathcal{O}(f)$ to avoid reintroducing fine-tuning problems the
scale $f$ needs to be around $1$ TeV. Little Higgs theories become
strongly coupled around a scale $\Lambda_{UV}\sim 4\pi f$ and need
to be UV completed at this scale.  I have listed the models in
existence thus far in Table~\ref{models} to show the economy of
the models, and what their generic features are.

\section{Electroweak Constraints}
I will now briefly discuss how we computed the Electroweak~(EW)
constraints on various Little Higgs models~\cite{meade1,meade2}.
For the Little Higgs models we looked at we treated them as an
effective field theory and integrated out the heavy particles.  We
then computed the tree level corrections(for the most part), to EW
observables.  A global fit was then performed to find a bound on
the scale $f$. If $f$ was required to be higher than $\sim 1$~TeV
then the Little Higgs model in question still required a certain
degree of fine tuning of the Higgs mass.

\begin{figure}[ht]
\centerline{\epsfysize=1.9cm\epsfbox{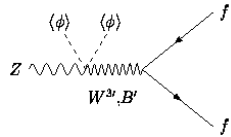}} \caption{Shift
in $\Gamma_z$ from coupling SM fermions to heavy gauge bosons.
\label{zwidth}}
\end{figure}

\par A question one might ask is why would constraints from
EW precision data be generically large?  A simple example can
illuminate this possibility, take for instance the generic feature
of heavy gauge bosons.  If the SM fermions couple to the heavy
gauge bosons it can cause a modification of the coupling of a Z to
two fermions as shown in Figure~\ref{zwidth}.  If one assumes the
mass of the heavy gauge bosons $W^{3'},B'$ to be around $f$ and
$c$ parameterizes the strength of the coupling between the heavy
and light fields then it is easy to express the shift from the SM
value as
\begin{equation}
\frac{\delta \Gamma_Z}{\Gamma_Z}\sim 1+c\frac{v^2}{f^2},
\end{equation}
where $v$ is the VEV of the Higgs field.  Since $\Gamma_Z$ is
measured extremely well, it is simple to calculate that if $c\sim
1$ then the EW bound on $f$ is
\begin{equation}
f>5.13\,\mathrm{TeV}\;\;\mathrm{to\;95\%\; C.L.}.
\end{equation}
This bound does not mean that in all Little Higgs model
$f>5.13\mathrm{TeV}$ since it is only one shift in the EW
precision data and the coupling $c$ was artificially set to $1$.
However, without doing precision EW fits of the various models
there is no reason a priori to believe that $f$ is naturally
around $1$~TeV. We analyze several models and their
variations\cite{littlest,anti,simple} in our
papers~\cite{meade1,meade2} and we find for generic regions of
parameter space the bound on $f$ is above $1$~TeV which implies a
certain degree of fine tuning.  In most models we find some range
of couplings, or a modification such that $f$ can be a TeV. The
biggest dangers for getting a large $f$ are from mixing between
heavy and SM gauge bosons, coupling of heavy $U(1)$ gauge bosons
to light SM fermions, Higgs triplet VEV's (all of these may be
sources of custodial $SU(2)$ violation) as well as new four-Fermi
operators that are introduced or very light new $U(1)$ gauge
bosons. Nevertheless since in most all models there exists a range
of parameter space such that $f\sim 1$~TeV it is ultimately
dependent on the UV completion of the model to tell us if that
range is natural.

\section*{Acknowledgments}
I wish to thank Csaba Cs\'{a}ki, Jay Hubisz, Graham Kribs, and
John Terning with whom I collaborated on the papers that led to
this talk.  I wish to thank the Graduate School of Cornell
University for partially supporting my travel to this conference.
This work was supported in part by the National Science Foundation
under Grant PHY/0139738.



\begin{thebibliography}{0}
\bibitem{georgi}
H.~Georgi and A.~Pais,
Phys.\ Rev.\ D {\bf 10}, 539 (1974).
H.~Georgi and A.~Pais,
Phys.\ Rev.\ D {\bf 12}, 508 (1975).

\bibitem{cohen}
N.~Arkani-Hamed, A.~G.~Cohen and H.~Georgi,
Phys.\ Lett.\ B {\bf 513}, 232 (2001).

\bibitem{littlest}
N.~Arkani-Hamed, A.~G.~Cohen, E.~Katz and A.~E.~Nelson,
JHEP {\bf 0207}, 034 (2002).

\bibitem{moose}
N.~Arkani-Hamed, A.~G.~Cohen, E.~Katz, A.~E.~Nelson, T.~Gregoire and J.~G.~Wacker,
JHEP {\bf 0208}, 021 (2002).

\bibitem{anti}
I.~Low, W.~Skiba and D.~Smith,
Phys.\ Rev.\ D {\bf 66}, 072001 (2002).

\bibitem{simple}
D.~E.~Kaplan and M.~Schmaltz,
JHEP {\bf 0310}, 039 (2003).

\bibitem{custmoose}
S.~Chang and J.~G.~Wacker,
arXiv:hep-ph/0303001.

\bibitem{simple2}
W.~Skiba and J.~Terning,
Phys.\ Rev.\ D {\bf 68}, 075001 (2003).

\bibitem{cust2}
S.~Chang,
JHEP {\bf 0312}, 057 (2003).


\bibitem{meade1}
C.~Csaki, J.~Hubisz, G.~D.~Kribs, P.~Meade and J.~Terning,
Phys.\ Rev.\ D {\bf 67}, 115002 (2003).

\bibitem{meade2}
C.~Csaki, J.~Hubisz, G.~D.~Kribs, P.~Meade and J.~Terning,
Phys.\ Rev.\ D {\bf 68}, 035009 (2003).



\end{thebibliography}
\end{document}